\documentclass[twocolumn, tighten, times]{aastex63}
\usepackage[utf8]{inputenc}
\usepackage{bm}
\usepackage{tabularx}
\usepackage{braket}
\usepackage{multirow}
\usepackage{xcolor}
\usepackage{pifont}
\usepackage{enumitem}
\usepackage{amsmath}
\usepackage{soul}
\usepackage{diagbox}
\usepackage{makecell}

\usepackage{lineno}

\shortauthors{ Hu }

\graphicspath{{./}{figures/}}

\begin{document}

\title{Exploring the possible evolution \textbf{of} the mass density power-law index of strong gravitational lenses with a model-independent method
}

\author[0000-0002-4797-4107]{Jian Hu}\thanks{E-mail:dg1626002@smail.nju.edu.cn}
\affiliation{School of Engineering, Dali University, Dali 671003, China}

\begin{abstract}
In this work, we adopt a cosmological model-independent approach for the first time to test the question of whether the mass density power-law index($\gamma$) of the strong gravitational lensing system(SGLS) evolves with redshift, and the JLA SNe Ia sample and the quasar sample from Risaliti \& Lusso (2019) are used to provide the luminosity distances to be calibrated. Our work is based on the flat universe assumption and the cosmic distance duality relation. A reliable data-matching method is used to pair SGLS-SNe and SGLS-quasar. By using the maximum likelihood method to constrain the luminosity distance and $\gamma$ index, we obtain the likelihood function values for the evolved and non-evolved cases, and then use \textbf{the Akaike weights and the BIC selection weights} to compare the advantages and disadvantages of these two cases. \textbf{We find that the $\gamma$ index is slightly more likely to be a non-evolutionary model for $\gamma=2$ in the case of the currently used samples with low redshift ($z_l<\sim$0.66).} \textbf{With Akaike weights}, the relative probability is 66.3\% versus 33.7\% and 69.9\% versus 30.1\% for the SGLS+SNe Ia sample and SGLS+quasar sample, respectively, and \textbf{with BIC selection weights}, the relative probability is 87.4\% versus 12.6\% and 52.0\% versus 48.0\% for the two samples. In the evolving case for the relatively low redshift lens (SGLS+SNe Ia), with redshift 0.0625 to 0.659, $\gamma= 2.058^{+0.041}_{-0.040}-0.136^{+0.163}_{-0.165}z$. At high redshift (SGLS+quasar ), with redshift 0.0625 to 1.004, $\gamma= 2.051^{+0.076}_{-0.077}-0.171^{+0.214}_{-0.196}z$. Although not the more likely model, this evolved $\gamma$ case also fits the data well, with a negative and mild evolution for both low and high redshift samples.
\end{abstract}


\section{Introduction} \label{sec:intro}
The Gravitational lensing system is a celestial phenomenon predicted by Einstein's general relativity. When the light from a distant object passes through a massive object, the light from the distant object will be deflected by the gravity generated by the object. Due to the different celestial bodies acting as lenses, gravitational lensing can be divided into strong gravitational lensing (SGLS), weak gravitational lensing, and gravitational microlensing. The first two phenomena usually consist of galaxies and galaxy clusters acting as lenses. Distortions in shape are easier to see with SGLS than with weak gravitational lensing. Since the discovery of the strong gravitational lensing of SGLS Q0957 + 561 by \cite{wal79}, this observational effect has gradually provided a new approach to observational cosmology. Nowadays, the SGLS is becoming more and more important in astronomical research, in particular, in cosmological models such as \cite{yuwa15,mel15,mel18,tu19}, dark energy fields such as \cite{zhu00,cha04,cao12,cao15,ama19},  cosmic distance duality relation (CDDR), $D_LD_{A}^{-1}(1+z)^{-2}=1$, the CDDR relates luminosity distances($D_L$) to angular diameter distances($D_A$) \citep{lia16,zho21,lim21}.

From observations of the SGLS, the radius of the Einstein ring and the velocity dispersion of the stars around the galaxy acting as a lens can be obtained. The radius of the Einstein ring is proportional to the square of the stellar velocity dispersion multiplied by the ratio of angular diameter distance (ADD). The ratio of angular diameter distance can be expressed as a ratio of the diameter distances($D_A$) between the lens and source and the observer and source.
A more precise relation depends on the adopted model of the lensing galaxy mass distribution. The usual model assumes a spherically symmetric power-law mass distribution ($\rho\propto r^{-\gamma}$) for the lensing galaxy mass fraction. More specifically, a power-law index of 2 gives the singular isothermal spherical profile (SIS model). The SIS model of the SGLS has been widely used for cosmological studies. For example, the CDDR \citep{hba16,rua18}, cosmic opacity\citep{ma19}, cosmological curvature\citep{ras15}. 

However, some researchers have found that the slope of an individual galaxy's density profile has a substantial impact on the SIS mode, causing non-negligible scattering \citep{koo05,koo09,son13,der21}. Therefore the determination of the lensing index $\gamma$ and whether it evolves with redshift becomes a key issue in cosmological studies, and overly simple models may introduce some systematic errors. \cite{hol17} used Type Ia Supernovae (SNe Ia)  and gamma-ray burst data (GRBs) to calibrate the distance modulus($D_L$) and study the evolution of the $\gamma$ index for a sample of 118 SGLS from \cite{cao15}, and they found that $\gamma =2$, which is not evolving with redshift within 1$\sigma$ confidence, but for different evolution parameter models $\gamma$ has a weak negative correlation with redshift. \cite{cui17} considered a spherically symmetric power-law mass distribution $\rho_{tot} (r)\sim r^{-\gamma}$ model using SNe Ia, Cosmic microwave background (CMB), baryon acoustic oscillation (BAO), and $H(z)$ data with multiple cosmological models to jointly constrain the cosmological model parameters and the $\gamma$. Their results show that the constraints on $\gamma$ are mainly determined by SGLS observations and are likely not driven by the choice of cosmological model, and find that $\gamma$ does not evolve with redshift. However, as their study has still made use of a specific cosmological model, it is necessary to test the evolution of the gamma index with a model-independent approach.

In this work, we adopt a cosmological model-independent approach to test whether $\gamma$ evolves with redshift and evaluated the advantage of both cases using the Akaike information criterion(AIC) and Bayesian Information Criterions(BIC). The luminosity distance is to be determined by matching SNe Ia or quasar observations near the redshift of the lens and the source of SGLS. To fit the SNe Ia light curve parameters or the related parameters of the quasars' optical-UV luminosity relation ($L_{X}-L_{UV}$ relation) and the $\gamma$ parameter and avoid the model dependence problem, the model of SGLS is used instead of a particular cosmological model. Note that the work of \cite{hol17}, although model-independent in its fit, directly adopts the luminosity distances of JLA SNe Ia  and GRB sample, which requires the assumption of a special cosmological model, i.e. their work uses data that are already model-dependent. Only our work is the first to use a model-independent approach to fit these parameters simultaneously. The CDDR and a flat universe assumption are also adopted to relate the luminosity distance to the angular diameter distance.

This paper is organized as follows. In Sec.2, we introduce the data set we use and the method of data pairing. In Sec.3, we describe the method of statistical analysis and the numerical results. The Conclusions and discussions are given in Sec.4.
\section{Data}\label{sec:Data}
In this section, we present the data used in this work, which include the SGLS sample, the SNe Ia sample, and the quasar sample. We also introduce the method we use of data pairing that we use to obtain two types of distance at the same redshift. 

\subsection{The strong gravitational lensing system sample}\label{sec:SGLS}
In this part, we provide the sample employed and a brief overview of the SGLS in this study. 

The sample of SGLS we used was from \cite{lea18}. This sample includes 158 SGLS, which are compilations from \cite{cao15} and \cite{shu17}. The redshifts of these SGLS are all determined spectroscopically, and the range of the lens redshifts is from 0.01 to 1.4. The angular radius of the strong gravitational lens for this sample ranges from $0.36"\sim 2.55"$. The velocity dispersion of this sample ranges from 98 km $ s^{-1}\sim 396$ km $s^{-1}$. After pairwise operations and elimination of non-physical cases, we selected 92 and 88 pairs for the SNeIa +SGLS and the quasar +SGLS samples, respectively. The lens redshift range of the two samples is 0.0625 to 0.659 and 0.0625 to 1.004, respectively.

SGLS is a potent astrophysical mechanism for investigating gravity and cosmology hypotheses. SGLS can be observed when the source, lens, and observer of the SGLS are sequentially aligned in an almost straight line. The arcs, multiple images, and even Einstein rings can usually be observed. In a cosmological background, it is usually a galaxy or quasar that acts as the source(s) and a galaxy or galaxy cluster that acts as the lens(l). The model of the mass distribution within the lens is usually related to the parameters of the lens system, considering the general case \citep{cao15}, the model of the lens of an SGLS with a spherically symmetric power-law mass distribution $\rho\propto r^{-\gamma}$, the ADD at this point can be written as
\begin{equation}\label{001}
	R^{A}_{len}(z_l,z_s)=\frac{D_{A_{ls}}(z_l,z_s)} {D_{A_{s}}(0,z_s)}=\frac{\theta_Ec^2}{4\pi\sigma_{ap}^2 f(\gamma)}\bigg({\theta_{ap}\over \theta_E}\bigg)^{2-\gamma},
\end{equation} 
where $c$ is the light speed, and $z_l$ and $z_s$ are the redshift of the lens and the source, respectively. $D_{A_{ls}}(z_l,z_s)$ and $D_{As}(0,z_s)$ are the angular diameter distances between the lens to the source and the source to the observer, respectively, and $\theta_{E}$ is the observed angular Einstein radius.
$\sigma_{ap}$ represents the velocity dispersion in an aperture with size $\theta_{ap}$, $\gamma$ represents the mass density power-law index. The function $f(\gamma)$ can be written by
\begin{equation}\label{002}
	f(\gamma)=-\frac{(5-2\gamma)(1-\gamma)}{\sqrt{\pi}(3-\gamma)}\frac{\Gamma(\gamma-1)}{\Gamma(\gamma-3/2)}\bigg[\frac{\Gamma(\gamma/2-1/2)}{\Gamma(\gamma/2)}\bigg]^2\,
\end{equation}
where $\Gamma(x)$ indicates the gamma function.

The deviation of $R^{A}(z_l,z_s)$ can be written by
\begin{equation}\label{003}
	\Delta{R^A}_{len}(z_l,z_s)=R^A(z_l,z_s)\sqrt{(4\delta_{\sigma_{ap}})^2+(1-\gamma)^2(\delta_{\theta_E})^2}.
\end{equation}
$\delta_{\sigma_{ap}}$ and $\delta_{\theta_E}$ are the fractional uncertainty of $\sigma_{ap}$ and $\theta_E$, respectively. By the work of SLACS team, we set the fractional uncertainty of the Einstein radius at 5\% .

By using the equation,
\begin{equation}\label{004}
	D_A(z)=\frac{D_C(z)}{1+z}=\frac{H_0d(z)}{c(1+z)},
\end{equation}
$D_C$ is the The comoving distance. $D_{A_{ls}}(z_l,z_s)$ can be expressed by
\begin{equation}\label{005}
	D_{A_{ls}}(z_l,z_s)=\frac{D_{Cls}(z_l,z_s)}{1+z_s},
\end{equation}

and $D_{A_{s}}(0,z_s)$ is expressed by
\begin{equation}\label{006}
	D_{A_{s}}(0,z_s)=\frac{D_{Cs}(z_l,z_s)}{1+z_s}=\frac{D_{C}(z_s)}{1+z_s}.
\end{equation}
In a flat universe, the dimensionless distance obeys the following relationship:
\begin{equation}\label{007}
	d(z_l,z_s)=d(z_s)-d(z_l).
\end{equation}
use the equation(\ref{007}), $R_{obs}^A(z_l,z_s)$ can be written as
\begin{equation}\label{008}
	R_{obs}^A(z_l,z_s)=1-\frac{D_C(z_l)}{D_C(z_s)}=1-\frac{(1+z_l)D_A(z_l)}{(1+z_s)D_A(z_s)}.
\end{equation}

Combining the CDDR, which can be expressed by 
\begin{equation}\label{009}
	\frac{D_L}{D_A}(1+z)^{-2}=1,
\end{equation}
the equation(\ref{008}) can also be written as
\begin{equation}\label{010}
	R_{obs}^A(z_l,z_s)=1-\frac{d(z_l)}{d(z_s)}=1-\frac{(1+z_s)D_L(z_l)}{(1+z_l)D_L(z_s)}.
\end{equation}
The subscript "obs" refers to the observation objects that provide luminosity distance, and in this paper refers to SNe and quasar.

	
In a nonflat space, the expression of $R_{obs}^A(z_l,z_s)$ is more complicated; one can refer to \cite{ras15}. Fortunately, most cosmological tests today support a flat cosmic space\citep{planck}. In this work, we investigate  the possible evolution of the mass density power-law index in flat space-time.

\subsection{The JLA SNe Ia Sample}\label{sec:The Pantheon SNe Ia Sample}
We use the joint light-curve (JLA) sample from \cite{bet14}, which includes 740 SNe Ia in the redshift range $0.01\leq z\leq 1.30$. When a specific cosmology model is used to fit the light curve parameters of the sample, the distance modulus can be calculated by these parameters, which assume that the SNe Ia have the same color, shape, and galactic environment have the same intrinsic brightness for all redshifts \cite{bet14}. This assumption may be quantified using an experimental linear relation, which results in a standard distance written as:

\begin{equation}\label{011}
	\mu=m_B^ \star-(M_B-\alpha \times X_1+\beta \times C),
\end{equation}
where "$m^{\star}_B$" is the observed peak magnitude in the B band of the rest- frame and "$\alpha$", "$\beta$", and "$M_B$" are nuisance parameters in the distance calculation. A simple step function relates the absolute magnitude to the stellar mass ($M_{stellar}$) of the host star by 

\begin{equation}\label{012}
	{M_B} = \left\{ \begin{array}{l}
		M_B^1\qquad \qquad {\rm{if}} ~~{M_{stellar}}{\rm{ <  1}}{{\rm{0}}^{10}}{M_ \odot }{\rm{      }},\\
		M_B^1 + {\Delta _M} \qquad {\rm{otherwise}}.
	\end{array} \right.
\end{equation}
The distance modulus is defined as
\begin{equation}\label{013}
\mu = 5 \log_{10}(\frac{D_L}{\rm Mpc})+25.
\end{equation}
Combining equations (\ref{011}) and (\ref{013}), the expression for the luminosity distance can be obtained, and the part $\frac{D_L(z_l)}{D_L(z_s)}$ of equation (\ref{010}) can be rewritten as
\begin{equation}\label{014}
\begin{split}
\log_{10}[\frac{D_L(z_l)}{D_L(z_s)}]=&0.2\{m_{B}(z_l)-m_{B}(z_s)+ \alpha[x(z_l)-x(z_s)]-\\
&\beta[c(z_l)-c(z_s)]+\Delta_M(z_l)-\Delta_M(z_s)\},
\end{split}
\end{equation}
the main part of $M_B$, $M_B^1$ in equation(\ref{012}), is offset, and only $\Delta _M$ remains. The deviation of $\frac{D_L(z_l)}{D_L(z_s)}$ can be written by
\begin{equation}\label{015}
\begin{split}
\Delta[\frac{D_L(z_l)}{D_L(z_s)}]_{(SNe)}=\frac{\ln{10}}{5}[\frac{D_L(z_l)}{D_L(z_s)}]\{\Delta^2_{m_{B}(z_l)}+\Delta^2_{m_{B}(z_s)}+ \\\alpha^2[\Delta^2_{x(z_l)}
+\Delta^2_{x(z_s)}]+\beta^2[\Delta^2_{c(z_l)}+\Delta^2_{c(z_s)}]+\Delta^2_{B(z_l)}+\Delta^2_B(z_s)+\\
2\alpha(covmx(z_l) +  covmx(z_s) ) -2\beta (covmc(z_l) + covmc(z_s)) \\
-2\alpha\beta(covxc(z_l) +  covxc(z_s))  \}^{0.5},
\end{split}
\end{equation}
where $\Delta_{m_{B}}$, $\Delta_{x}$, and $\Delta_{c}$ are the deviation of visual magnitude $m_B$, the observation $x$, and $c$, and covmx, covmc,covxc are their covariance, respectively. $\Delta_{B}$ is the bias correction uncertainty. These quantities are all from \cite{bet14}.

For an evolutionary $\gamma$ in equation(\ref{001}), we take a simple parameterization method expressed as follows
\begin{equation}\label{016}
	\gamma_0=\gamma_1+\gamma_2 z_l,
\end{equation}


\subsection{The quasar Sample}\label{sec:quasar}
In this article, the quasar sample we used from \cite{risa19}. This sample compiles 1598 quasars with the redshift range of $0.1 \leq z \leq 5.1$.

According to the accepted quasar accretion model, the matter around the accretion disk is partially converted to ultraviolet (UV) radiation by the gravitational force of the active galactic nucleus \citep{sha73}, and some of these ultraviolet photons undergo inverse Compton scattering with hot relativistic electrons from around the accretion disk, resulting in x-ray radiation. A nonlinear relationship between the luminosity of these two components, the $L_X-L_{UV}$ relation, has been found for more than 30 years \citep{avn86}, and this relationship can be written as
\begin{equation}\label{017}
	\log_{10}{L_X}=\gamma\log_{10}{L_{UV}}+\beta,
\end{equation}
where $L_X$(X denotes X-ray band ) and $L_{UV}$(UV denotes UV bands) are the monochromatic luminosities in the rest-frame at 2 keV and 2500 $\dot{A}$ , the slope $\gamma$ is a free parameter with$\sim$ 0.6 \citep{risa19}; and $\beta$ is a normalization constant of this relation. For several different samples from previous studies \citep{jus07,lus10,you10}, the observed dispersion $\delta$ is on the order of 0.35$\sim$0.40.
According to the relationship between luminosity and luminosity distance and fluxes, the equation(\ref{016}) can be expressed by
\begin{equation}\label{018}
	\log_{10}{F_X}=\beta^{\prime}+\gamma\log_{10}{F_{UV}}+2(\gamma-1)\log_{10}{D_L},
\end{equation}
where $\beta^{\prime}$ is also a constant that includes both the slope $\gamma$ and the constant $\beta$ in equation(\ref{017}), which can be written by
\begin{equation}\label{019}
	\beta^{\prime}=\beta+(\gamma-1)\log_{10}{4\pi}.
\end{equation}

According to equation (\ref{018}), the expression for the luminosity distance also can be obtained, and the part $\frac{D_L(z_l)}{D_L(z_s)}$ of equation (\ref{010}) can be rewritten as
\begin{equation}\label{020}
	\begin{split}
\log_{10}[\frac{D_L(z_l)}{D_L(z_s)}]=\frac{1}{2(\gamma-1)}[\log_{10}F_X(z_l)-\log_{10}F_X(z_s)\\
-\gamma(\log_{10}F_{UV}(z_l))-\log_{10}F_{UV}(z_s)],
	\end{split}
\end{equation}
The parameter $\beta^{\prime}$ is counteracted, and the deviation of $\frac{D_L(z_l)}{D_L(z_s)}$ can be expressed by
\begin{equation}\label{021}
	\begin{split}
		\Delta[\frac{D_L(z_l)}{D_L(z_s)}]_{(quasar)}=\frac{\ln{10}}{2(\gamma-1)}[\frac{D_L(z_l)}{D_L(z_s)}]\\
		\{\Delta^2(\log_{10}F_X(z_l))+\Delta^2(\log_{10}F_X(z_s))+2\delta^2  \}^{0.5}
	\end{split}
\end{equation}
where $\Delta\log_{10}F_X$ is the observation uncertainty of $\log_{10}F_X$, which can be found in \cite{risa19} and the observation uncertainty of $\log_{10}F_{UV}$ is considered to be small in comparison to the $\log_{10}F_X$, and is so neglected in this work. $\delta$ is a free parameter, representing the uncertainty of the $L_X-L_{UV}$ relation.

\subsection{Data pairing}\label{sec:data pairing}
In this section, we will provide a brief introduction to the technique of data pairing. To explore the mass density power-law index, the angular diameter distances in equation (\ref{001}) should be given. Using the CDDR, the luminosity distance of SNe Ia and quasar can provide an angular diameter distance at the redshift of the SNe or the quasar. According to equation(\ref{010}), the distance we actually need to use is the dimensionless distance. If the redshifts of the source and lens of SGLS are close enough to the object giving the dimensionless distance (SNe Ia or quasar), we can utilize the dimensionless distance of other objects rather than the dimensionless distance of the source and lens.. 

We know that the closer the redshift, the smaller the distance between the two objects. Therefore, many researchers use a fixed redshift deviation to pair the data. For example, setting $\Delta z=0.005$ \citep{hol10,hol12,hba16,hu18}, and others set the other value, \cite{ghj12} set the value for $0.006$, and \cite{lia19} set it to 0.003. The advantage of this setup is that it is model independent on the matched data pair, but because redshift is not linear for cosmological distances (luminosity distance, dimensionless distance, and angular diameter distance), this makes it difficult to match the data at high redshifts, resulting in waste of data. Once matched, its distance deviation is a small value, while at low redshifts, according to such a $\Delta z$ to match its distance deviation is very large, and the distance deviation is not consistent.

\begin{figure}
	\centering
	\includegraphics[scale=0.6]{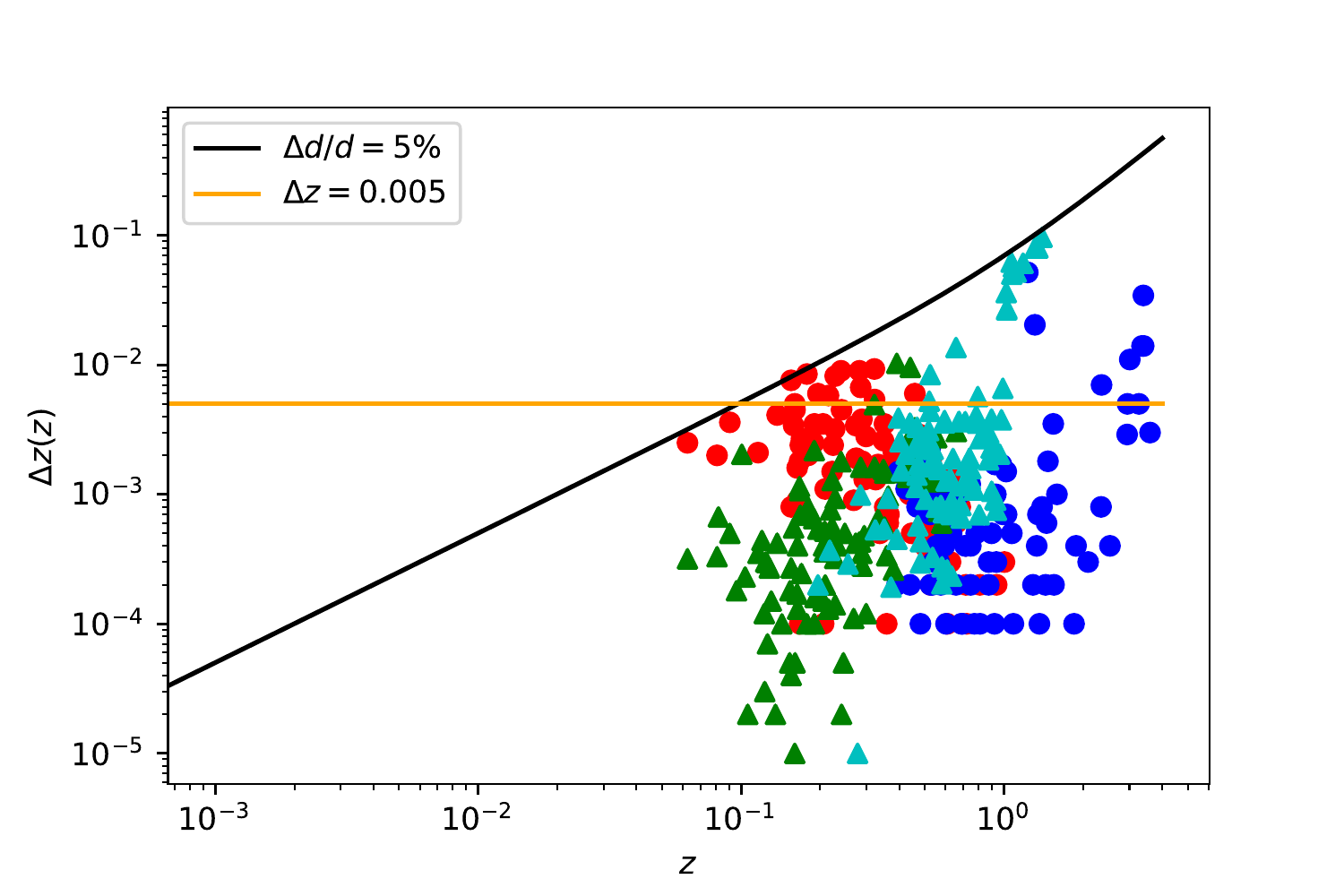}\\
	\caption{The black solid line  indicate the relation between the $\Delta z$ and $z$ in $\Lambda$CDM with $\Delta d/d=5\%$. The orange solid line indicates $\Delta z=0.005$. The green triangle represents the absolute difference between the lens redshift of a SGLS and the redshift of a pair of SNe Ia. The indigo triangle represents the absolute difference between the redshift of the source of a SGLS and that of a paired SNe Ia. The red dots represent the absolute difference between the redshift of the lens of the SGLS and that of the matched Quasar, and the Blue Dots represent the absolute difference between the redshift of the source of the SGLS and that of the matched Quasar.}
	\label{fig01}
\end{figure}

To solve these problems, we adopt the method of distance (dimensionless distance) deviation consistency, which was first proposed by our previous paper \citep{zho21}. We have refined the details of the calculation here. 
The idea is to set a fixed $\Delta d/d$, where $d$ is the dimensionless distance and $\Delta d$ as its deviation, where $d$ and $z$ must have a certain relationship. We used the flat $\Lambda$CDM model with setting $\Omega_m$ = 0.31 and $\Delta d/d=5\%$. The redshift $z$ and $\Delta z$ should conform to the following relationship
\begin{equation}\label{022}
\begin{split}
d^{\Lambda CDM}(z+\Delta z)/d^{\Lambda CDM}(z)-1=5\%,
\end{split}
\end{equation}
Where the expression for $d_c^{\Lambda CDM}$ is written as
\begin{equation}\label{023}
d^{\Lambda CDM}(z)= \int_{0}^{z}\frac{dz^{\prime}}{\sqrt{0.31(1+z^{\prime})^3+0.69}}
\end{equation}
The ordinary differential equation of $\Delta z$ for $z$ is given by the combination of equations (\ref{022}) and (\ref{023}), which can be written by
\begin{equation}\label{024}
	\frac{d\Delta z(z)}{dz}=1.05 \frac{\sqrt{0.31(1+z+\Delta z(z))^3+0.69}}{\sqrt{0.31(1+z)^3+0.69}}-1,
\end{equation}
with initial solution $\Delta z(0)=0$. The numerical solution of equation (\ref{024}) is shown in the solid black line portion of figure(\ref{fig01}). As a comparison, we give the line $\Delta z=0.005$, the solid orange line in figure (\ref{fig01}). For two data samples, for example, SNe and SGLS sample, because the selection method that satisfies the condition is not unique, to select the best sample, we calculate the sum of the redshift deviation of the two groups of data samples, and the best sample should satisfy the minimum value of $\sum\limits_{i = 1}^{2n} {(\Delta z_i^k} {)^2}$, where $n$ is the logarithm number of pairing, and $k$ is the $k$ alternative. Note that the pairing data used here is based on a special cosmological model, but this model provides only a maximum fixed deviation of $d$, which is not taken into account in the later calculations and is only used to rule out large $\Delta d$. 
The advantage of this is that the selection deviation at high and low redshifts is consistent, or less than a fixed value at the same time, and more data at high redshifts can be paired. For the SGLS subsamples of 128 and 740 supernovae, we get 92 sets of data with the $z_{l max}=0.659$ of the lens and $z_{s max}=1.396$ of the source for the SGLS sample. For the former and 1598 quasars, we get 109 sets, excluding the non-physical case, $d_{ls}/d_s > 1$ for $\gamma \sim0.6$, we got 88 sets of data with the $z_{l max}=1.004$ of the lens and $z_{s max}=3.595$ of the source  for the SGLS sample. According to figure (\ref{fig01}), we can see the superiority of this method. If we adopt the fixed $\Delta z$ method, we lose the high redshift information in the upper right region of the figure. And our method preserves them without increasing the statistical deviation.

\section{Method and results} \label{sec:Method and results}
In this section, we introduce the statistical method and show the result.

\subsection{Method of statistical analysis}\label{sec:method of statistical analysis}
We first present the fitted parameters and likelihood functions for the evolving and non-evolving cases, and then use the maximized likelihood function to obtain the best value and corresponding confidence interval for each parameter. The difference between the two cases is the $\gamma_2$ parameter in equation (\ref{016}). Taking $\gamma_2=0$ describes the non-evolutionary case, and vice versa describes the evolutionary case. For SGLS+SNe Ia sample, the fitted parameters $\theta$ are $\{\alpha, \beta, \Delta_M,\gamma_0\}$ and $\{\alpha, \beta, \Delta_M,\gamma_1, \gamma_2\}$ for these two cases, and $\{\gamma_0, \gamma, \delta\}$ and $\{\gamma_1, \gamma_2, \gamma, \delta\}$ for SGLS+quasar respectively. To determine the best-fitting values of each parameter, we use the method of maximum likelihood estimation. Substituting equation (\ref{014}) into equation (\ref{010}) and combining it with equation (\ref{001}), their likelihood functions can be written uniformly as
\begin{equation}\label{025}
	\begin{split}
		\ln(L(\theta|x))=-\frac{1}{2}\sum_{1}^{n}(\frac{(R^{A}_{len}(z_l,z_s)-R^{A}_{SNe}(z_l,z_s))^2}{(\Delta{R^A}_{len}(z_l,z_s)^2+(\Delta R^A_{obs}(z_l,z_s))^2)}\\
		+\ln(2\pi(\Delta{R^A}_{len}(z_l,z_s)^2+(\Delta R^A_{obs}(z_l,z_s))^2)).
	\end{split}
\end{equation}
Different $\theta$, used to describe the fitting parameters for the four cases above. The $x$ in equation(\ref*{025}) represents the observation data. To constrain the confidence intervals for $\theta$ in equation(\ref{025}), the Markov-Monte Carlo (MCMC) method is used, from the Python package (named emcee) by \cite{fore13}. We use the prior value of $\theta=(0.140,3.139,-0.060,2.000)$ or $\theta=(0.140,3.139,-0.060,2.000,0)$  from \cite{bet14} and an SIS model of SGLS hypothesis. The prior interval we adopt covers the prior values and assumes that these parameters are uniformly distributed. The deviation of the JLA sample of SNe Ia consists of two kinds, a statistical uncertainty, and a systematic uncertainty, and since we are using a subsample of this sample, we only use the statistical uncertainty. Then the uncertainty of $R_{obs}^A(z_l,z_s)$ from equation(\ref{010}) can be expressed by
\begin{equation}\label{026}
	\Delta R_{SNe}^A(z_l,z_s)=-\frac{(1+z_s)}{(1+z_l)}\Delta[\frac{D_L(z_l)}{D_L(z_s)}]_{(SNe)}.
\end{equation}

For the SGLS+quasar sample,  $\theta=(\gamma_0, \gamma, \delta)$ for a nonevolutionary case, and $\theta=(\gamma_1, \gamma_2, \gamma, \delta)$ for an evolution case. We use the prior value of $\theta=(2.000,0.600,0.150)$ or $\theta=(2.000,0,0.600,0.150)$  from \cite{risa19} and also an SIS model of the SGLS hypothesis. Similarly, uncertainty of $R_{obs}^A(z_l,z_s)$ from equation(\ref{010}) at this point can be expressed as

\begin{equation}\label{027}
		\Delta R_{quasar}^A(z_l,z_s)=-\frac{(1+z_s)}{(1+z_l)}\Delta[\frac{D_L(z_l)}{D_L(z_s)}]_{(quasar)}.
\end{equation}

To quantify the possibility of the evolution index of SGLS, we use the AIC and BIC. They are defined as
\begin{equation}\label{028}
	AIC=-2\ln(L)+2k,
\end{equation}
and
\begin{equation}\label{029}
	BIC=-2\ln(L)+k\ln(n).
\end{equation}
$k$ is the number of free parameters in the two cases, n is the number of the subsample in section(\ref{sec:Data}), and $L$ is the likelihood value in equation(\ref{025}) with the best $x$. \textbf{To make a scientific decision between the two models, we need to define $\Delta XIC$, where $\Delta XIC = XIC_2 - XIC_1$, where $XIC$ represents the two information criterion. Generally speaking, if $\Delta XIC$ is equal to or less than 2, the prospective model is significantly endorsed compared to the reference model. If $\Delta XIC$ falls between 4 and 7, it suggests that the particular model is comparatively less endorsed than the reference model. If $\Delta XIC$ is equal to or greater than 10, the candidate model has no empirical support \citep{nun17}. If $\Delta XIC$ cannot distinguish the merits of the models very well, we introduce the Akaike weights and the BIC selection weights to describe the relative likelihood of a model \citep{and04}. The probability of $model_{\alpha}$  over the $model_{other}$ is }

\begin{equation}\label{030}
	P(\alpha)=\frac{\exp{(-XIC_{\alpha}/2)}}{\exp{(-XIC_{other}/2)}+\exp{(-XIC_{\alpha}/2)}}.
\end{equation}

\subsection{Results}
\begin{figure}
	\includegraphics[scale=0.42]{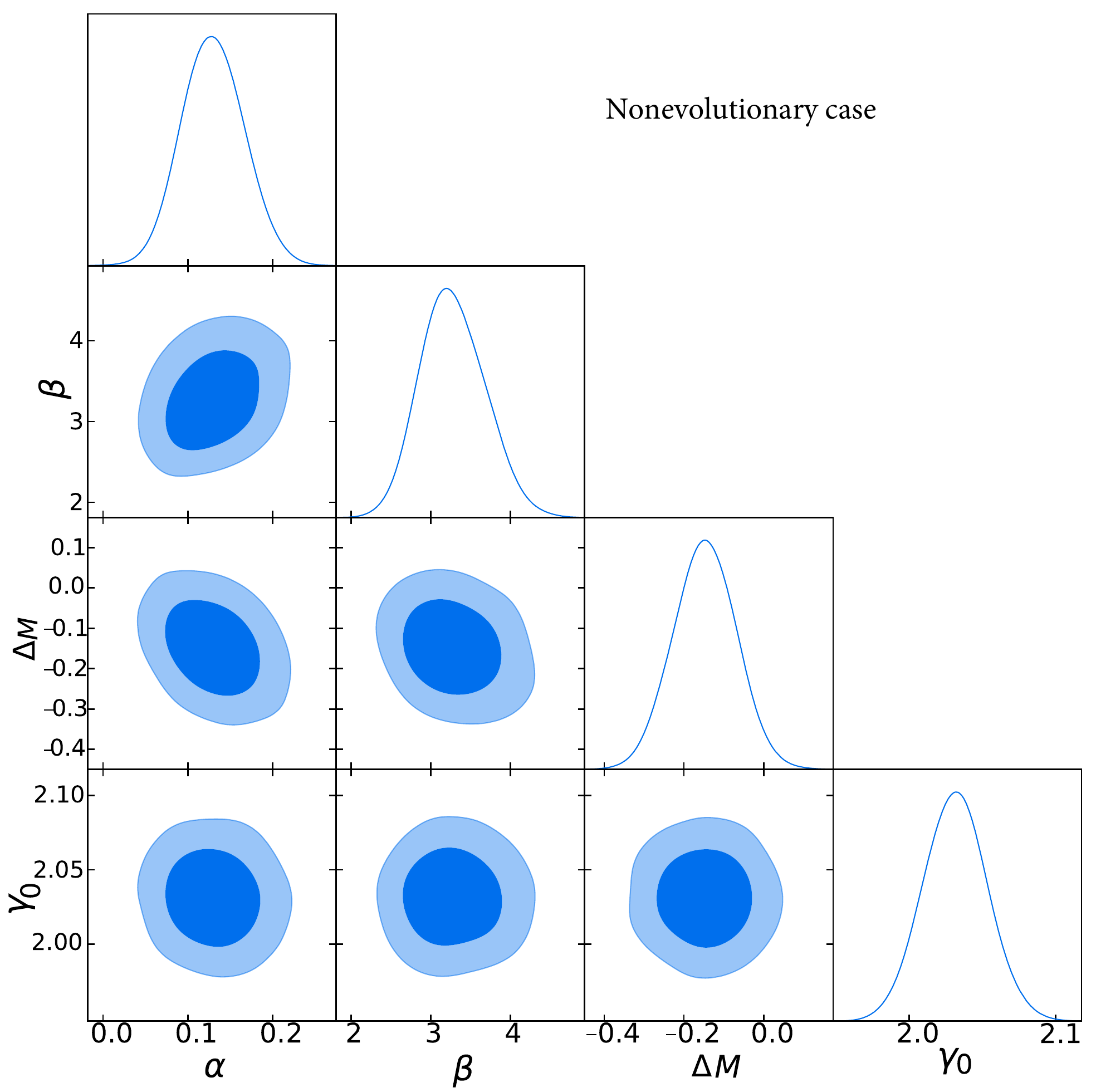}\\
	\caption{The 2D regions and 1D marginalized distributions with 1$\sigma$ and 2$\sigma$ contours for the parameters $\alpha$, $\beta$, $\Delta M$, and $\gamma_0$ using the SGLS+SNeIa sample.}
	\label{fig02}
\end{figure}

\begin{figure}
  \includegraphics[scale=0.38]{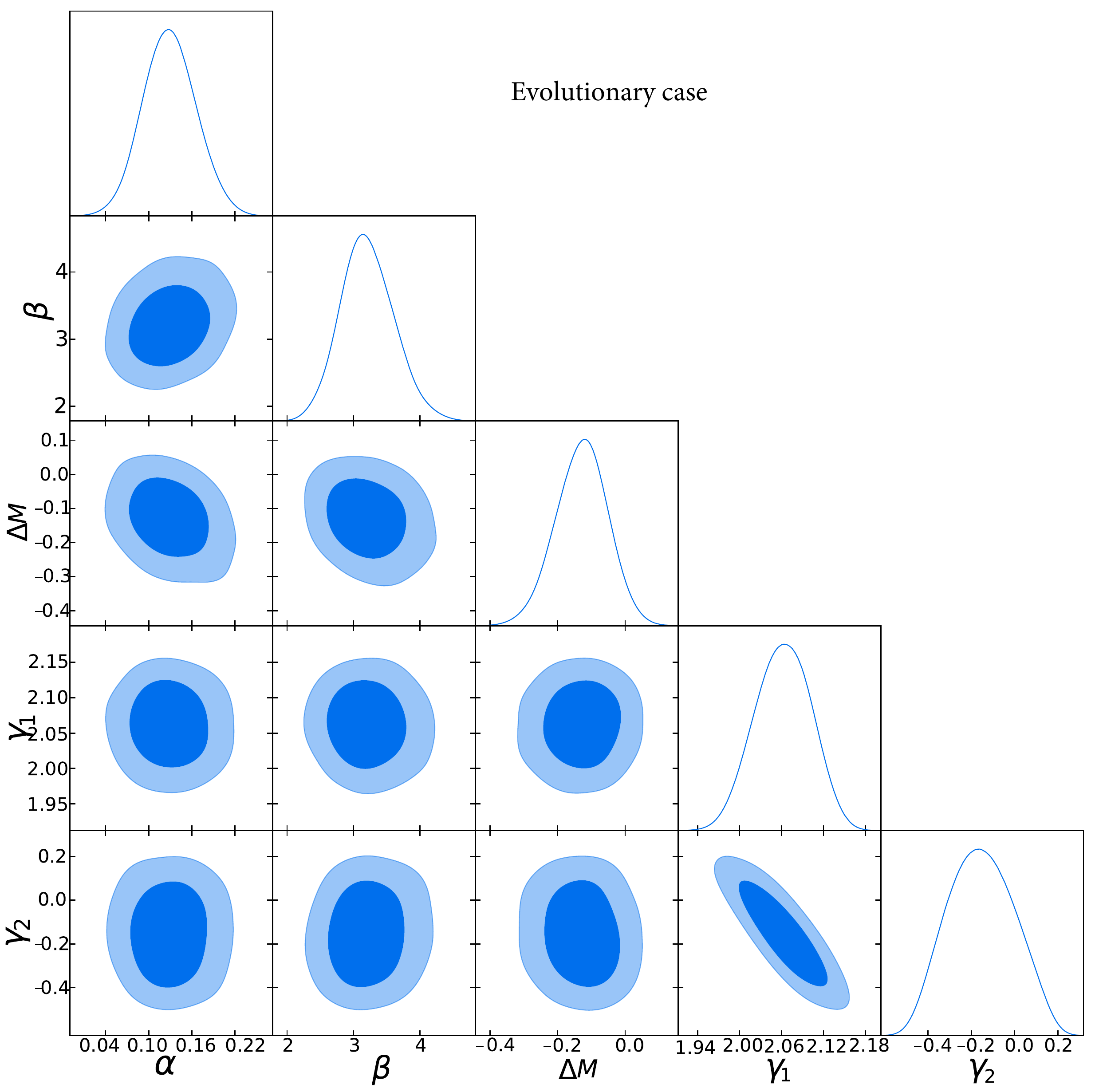}\\
  \caption{The 2D regions and 1D marginalized distributions with 1$\sigma$ and 2$\sigma$ contours for the parameters $\alpha$, $\beta$, $\Delta M$, $\gamma_1$ and $\gamma_2$ using the SGLS+SNeIa sample.}
  \label{fig03}
\end{figure}

The results are shown in Figure (\ref{fig02}), Figure (\ref{fig03}), and Table (\ref{tab1}) for the sample of SGLS+SNe Ia. From Table (\ref{tab1}), \textbf{we can see that the $\Delta AIC=1.353<2$ and $2<\Delta BIC=3.875<4$. According to the introduction of $\Delta XIC$ in Section (\ref{sec:method of statistical analysis}), $\Delta XIC$ is far from indicating that one model is better than another. If we use the Akaike weights and the BIC selection weights,} the nonevolutionary case of $\gamma$ \textbf{might} more consistent with the sample, with a relative probability of 66.9\% versus 33.7\% and 87.4\% versus 12.6\% for both the Akaike weights and the BIC selection weights, respectively. In the nonevolutionary case, The mass density power-law index $\gamma_0=2.032^{+0.021}_{-0.022}$ for 1$\sigma$ confidence level. In the evolutionary case, $\gamma_1=2.058^{+0.041}_{-0.040}$ and $\gamma_2=-0.136^{+0.163}_{-0.165}$ for 1$\sigma$ confidence level. The $\gamma_0$ parameter for the non-evolving case is thus close to 2, which approximates the SIS model well.

\begin{figure}
	\includegraphics[scale=0.58]{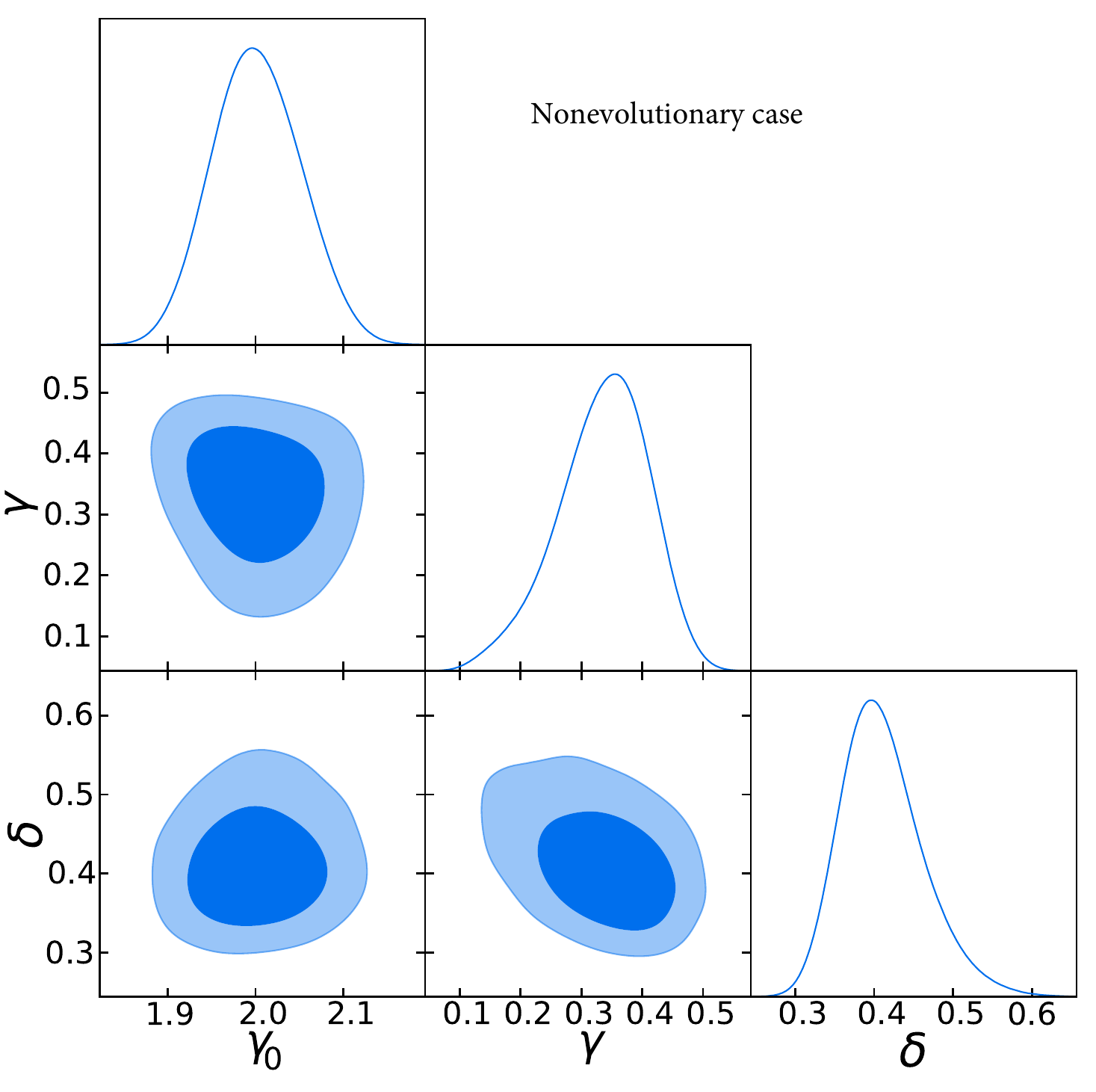}\\
	\caption{The 2D regions and 1D marginalized distributions with 1$\sigma$ and 2$\sigma$ contours for the parameters $\gamma_0$, $\gamma$, and $\delta$, using the SGLS+quasar sample.}
	\label{fig04}
\end{figure}

\begin{figure}
	\includegraphics[scale=0.42]{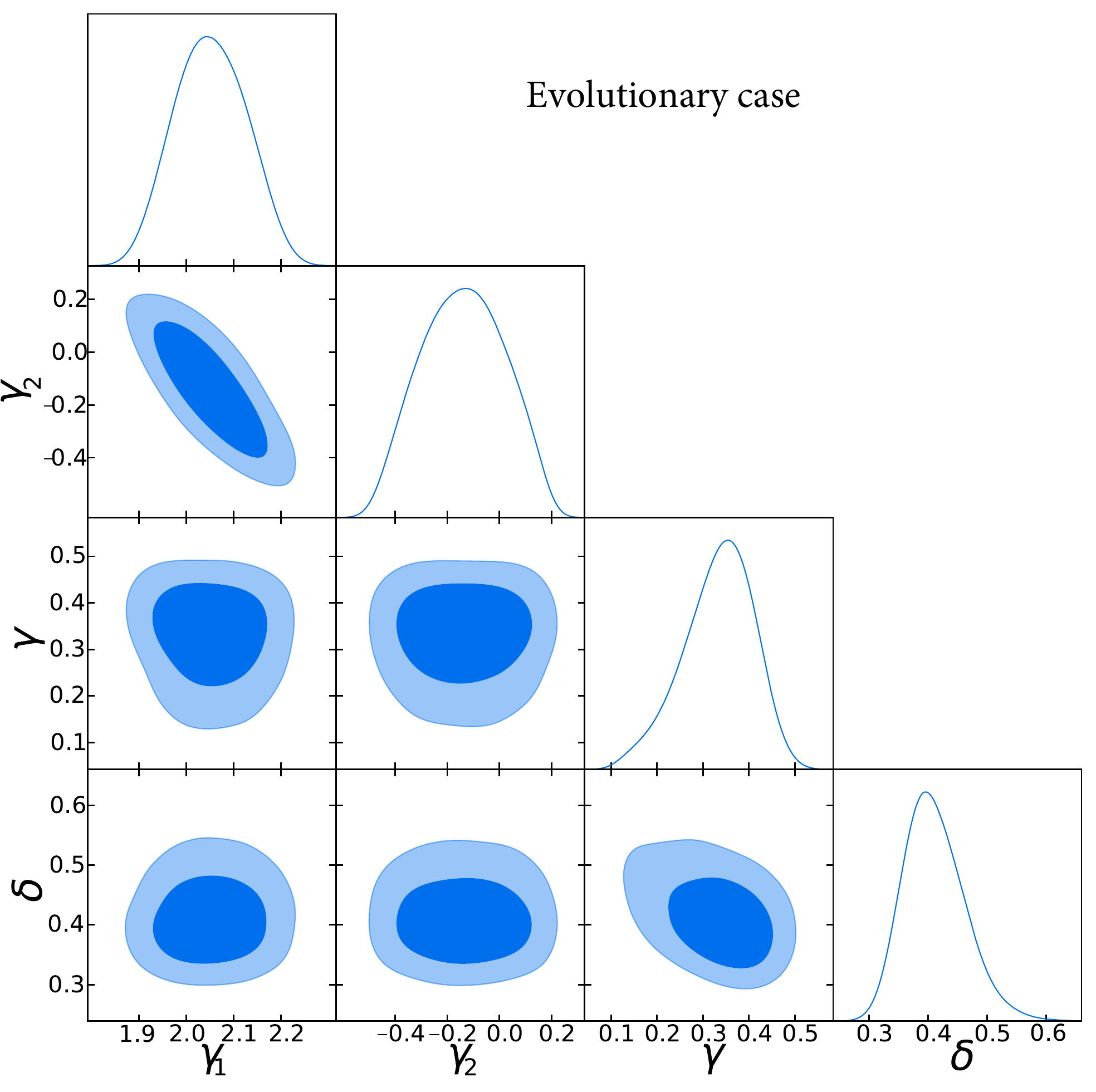}\\
	\caption{The 2D regions and 1D marginalized distributions with 1$\sigma$ and 2$\sigma$ contours for the parameters $\gamma_1$, $\gamma_2$,$\gamma$, and $\delta$, using the SGLS+quasar sample.}
	\label{fig05}
\end{figure}

\begin{figure}
	\includegraphics[scale=0.68]{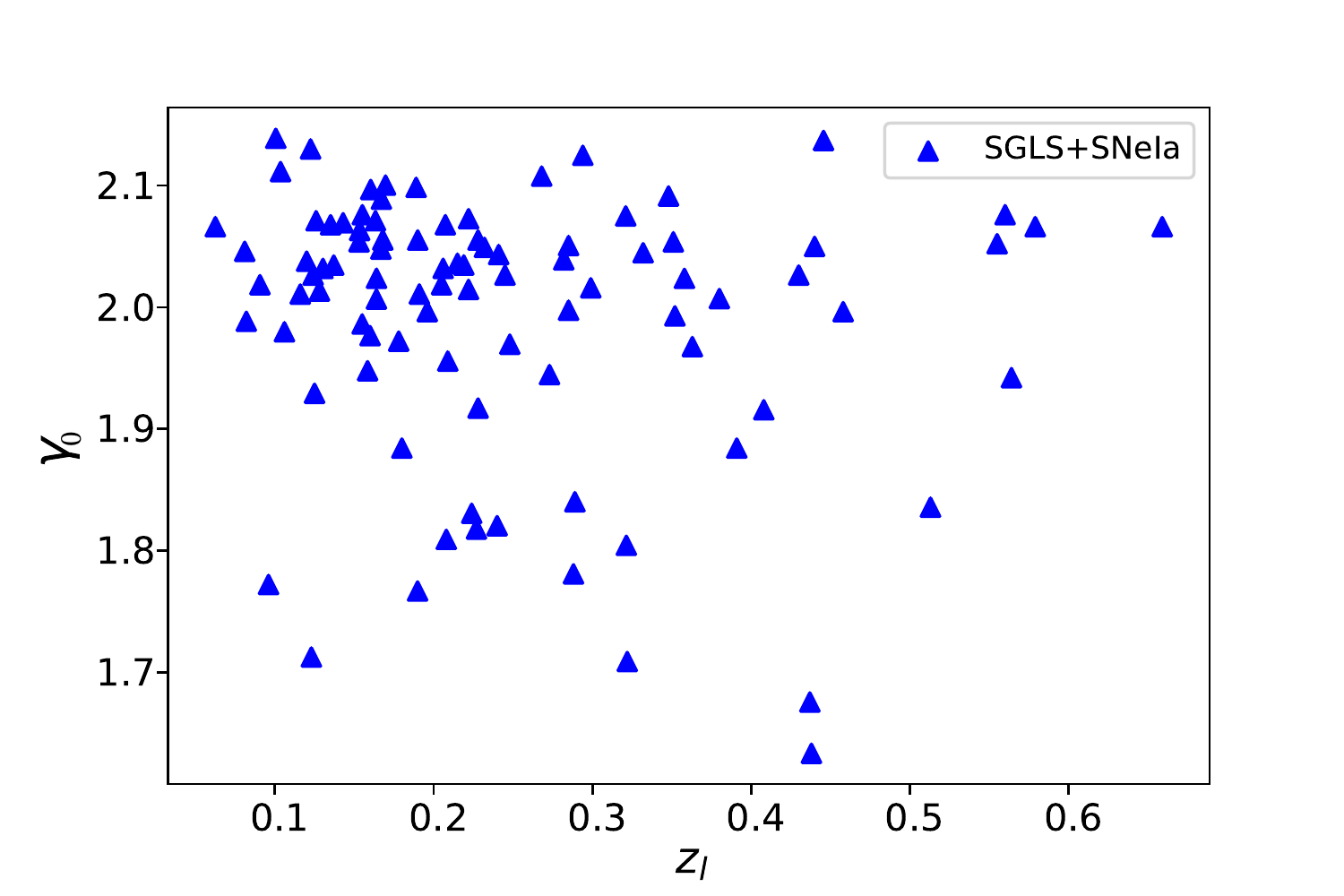}\\
	\caption{The variation of the parameter $\gamma_0$ (equation (\ref{016})) with $z_l$ when fixing $\alpha$, $\beta$, and $\Delta M$ as the best-fit value for the evolutionary case in Table(\ref{tab1}).}
	\label{fig06}
\end{figure}

\begin{figure}
	\includegraphics[scale=0.68]{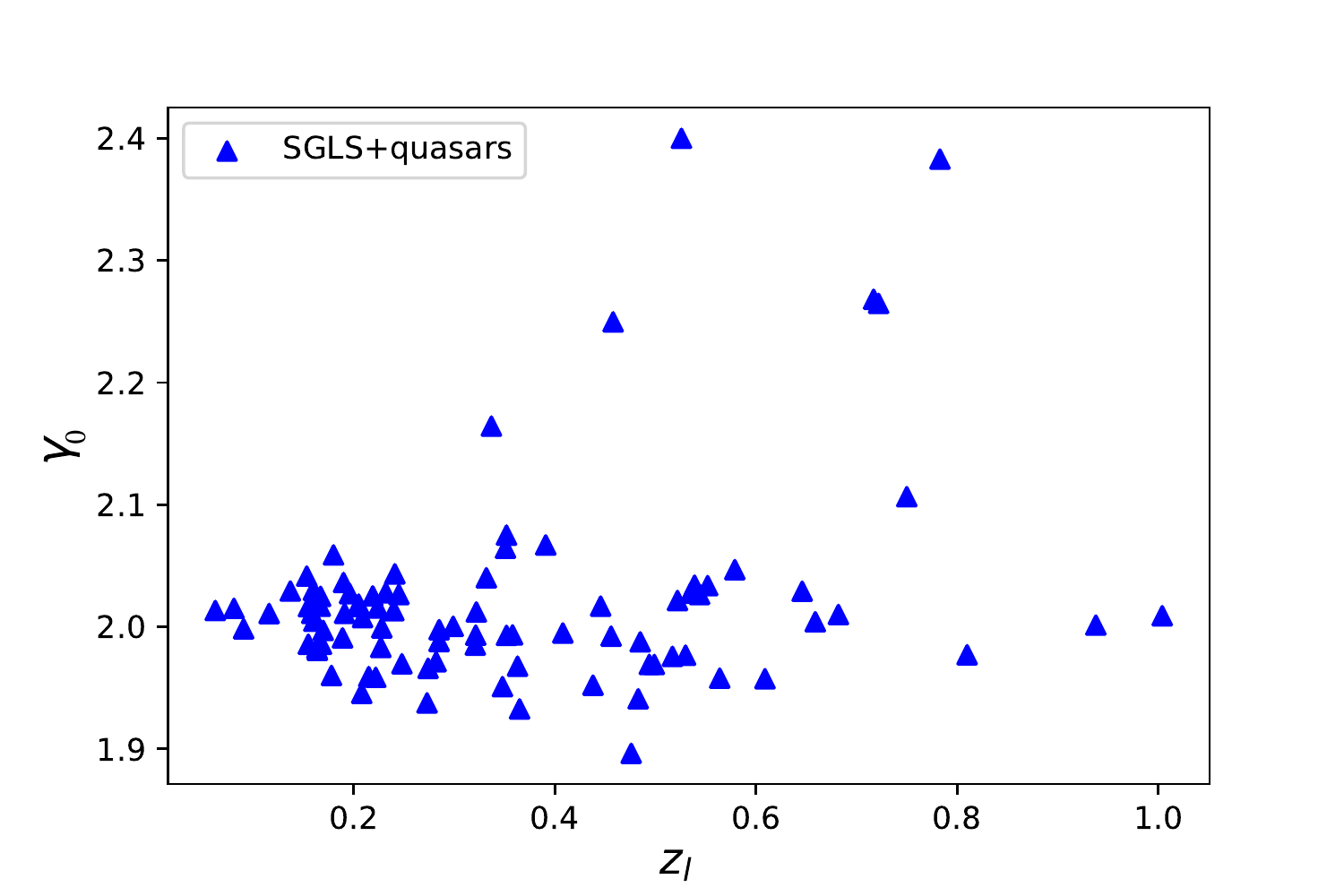}\\
	\caption{The variation of the parameter $\gamma_0$ (equation (\ref{016})) with $z_l$ when fixing $\gamma$ and $\delta$ as the best-fit value for the evolutionary case in Table(\ref{tab2}).}
	\label{fig07}
\end{figure}

The results for the sample of SGLS+quasar are shown in Figure (\ref{fig04}), Figure (\ref{fig05}), and Table (\ref{tab2}). \textbf{According to Table (\ref{tab2}), the results show that $\Delta AIC=1.687<2$ and $\Delta BIC=0.164<2$, $\Delta XIC$ still cannot exclude any one model. Using the Akaike weights and the BIC selection weights, we can see that the relative probability of 69.9\% versus 30.1\% and 52.0\% versus 48.0\% favor the nonevolutionary preferred case versus the evolutionary preferred case, respectively. This might indicate that the nonevolutionary case still outperforms the evolved case and that the weak advantage of the nonevolutionary case becomes even weaker with the inclusion of high redshift data ($0.66\sim1.0$).} The evolution of $\gamma_0$ with $z_l$ is shown in Figure (\ref{fig06}) and Figures (\ref{fig07}) for the SGLS+SNe Ia and SGLS+quasar samples, respectively, and when the other parameters are fixed to the best-fit values of Table (\ref{tab1}) and Table (\ref{tab2}). These two figures show that the evolution of $\gamma_0$ with $z_l$ is weak at relatively low redshifts ($<\sim0.66$), while at high redshifts ($0.66\sim1.0$) it may be difficult to describe with a simple linear model.

\subsection{Comparing results}
In this subsection, we will compare our data, methods, and results with those of previous work in various aspects, which will be shown in Table(\ref{tab3}). We compare only the more relevant parts of the work, i.e., the case where the $\gamma$ exponential evolution is assumed. Based on the fitted parameter, it is clear that the $\gamma_2$ is negative under the assumption of the evolutional case , that is, $\gamma$ decreases with increasing redshift, which is consistent with the previous works. However, considering the 1$\sigma$ range of the $\gamma_2$ parameter, a positive slope is also highly likely, which is still consistent with the non-evolutionary results, as is previous work. This may require a larger sample of data for further study.

\begin{table}
	\caption{Constraints on the light-curve parameters of SNe Ia and the mass density power-law index of the SGLS at the 1$\sigma$ confidence levels with the SGLS+SNeIa sample.} \scalebox{1.0}{
		\begin{tabular}{|c|l|l|}
			\hline
			parameters &             nonevolution  &evolution \\
			\hline
			$\gamma_0$  &  $2.032^{+0.021}_{-0.022}$ & $\ldots$ \\
			\hline
			$\gamma_1$  &        $\ldots$&        $2.058^{+0.041}_{-0.040}$\\
			\hline
			$\gamma_2$  &        $\ldots$   &        $-0.136^{+0.163}_{-0.165}$    \\
			\hline
			$\alpha$ &        $0.129^{+0.038}_{-0.036}$   &        $0.130^{+0.038}_{-0.038}$    \\
			\hline
			$\beta$ &        $3.255^{+0.440}_{-0.386}$ &        $3.230^{+0.405}_{-0.380}$ \\
			\hline
			$\Delta M$ &        $-0.147^{+0.078}_{-0.079}$ &        $-0.136^{+0.163}_{-0.165}$ \\
			\hline
			$\ln(L)$ &       40.713&        41.037      \\
			\hline
			AIC(P($\alpha$)) &        -73.426(66.3\%)  &          -72.073(33.7\%)           \\
			\hline
			$\Delta$AIC&    0&1.353\\
			\hline
			BIC(P($\alpha$)) &        -63.339(87.4\%) &          -59.464(12.6\%)           \\
			\hline
			$\Delta$BIC&    0&3.875\\
			\hline
	\end{tabular}}\label{tab1}
\end{table}

\begin{table}
	\caption{Constraints on the slope parameter of $L_X-L_{UV}$ relation of the quasar, and the mass density power-law index of the SGLS at the 1$\sigma$ confidence levels with the SGLS+quasar sample.} \scalebox{0.9}{
		\begin{tabular}{|c|l|l|}
			\hline
			parameters &             nonevolution  &evolution \\
			\hline
			$\gamma_0$  &  $2.000^{+0.053}_{-0.049}$ & $\ldots$ \\
			\hline
			$\gamma_1$  &        $\ldots$&        $2.051^{+0.076}_{-0.077}$\\
			\hline
			$\gamma_2$  &        $\ldots$   &        $-0.171^{+0.214}_{-0.196}$    \\
			\hline
			$\gamma$ &        $0.341^{+0.069}_{-0.082}$   &        $-0.345^{+0.067}_{-0.084}$    \\
			\hline
			$\delta$ &        $0.406^{+0.056}_{-0.045}$ &        $0.408^{+0.055}_{-0.047}$ \\

			\hline
			$\ln(L)$ &       -85.742&        -85.586      \\
			\hline
			AIC(P($\alpha$))  &        177.485(69.9\%) &          179.172(30.1\%)            \\
			\hline
			$\Delta$AIC&    0&1.687\\
			\hline
			BIC(P($\alpha$)) &        188.917(52.0\%) &          189.081((48.0\%))           \\
			\hline
			$\Delta$AIC&    0&0.164\\
			\hline
	\end{tabular}}\label{tab2}
\end{table}

\begin{table*}
	\caption{A comparison with previous related work.} \scalebox{0.8}{
		\begin{tabular}{|c|c|c|c|c|}
			\hline
			work &             data  & \makecell{selecting data\\method,(number)} & fiting method &results\\
			\hline
			\cite{cao15}  &118SGLS& $\ldots$ & \makecell{Minimize $\chi^2$ with \\XCDM model} & \makecell{$\gamma_1=2.06\pm0.09$\\ $\gamma_2=-0.09\pm 0.16$} \\
			\hline
			\cite{cui17}  & \makecell{118SGLS+\\JLA SNe Ia}       &        \ldots  & \makecell{Minimize $\chi^2$ with \\XCDM model}&\makecell{$\gamma_1=2.103^{+0.045}_{-0.054}$\\ $\gamma_2=-0.062^{+0.096}_{-0.105}$} \\
			\hline
			\cite{hol17}  &  \makecell{118SGLS+Union2.1 \\SNe Ia+GRB\\118SGLS+JLA \\SNe Ia+GRB}   &\makecell{$\Delta z \leq 0.006$, (92) \\$\Delta z \leq 0.006$, (87)}   & \makecell{Minimize $\chi^2$\\Implicit \\model dependency}&\makecell{$\gamma_1=2.04^{+0.08}_{-0.06}$\\ $\gamma_2=-0.085^{+0.21}_{-0.18}$\\ $\gamma_1=2.04^{+0.08}_{-0.06}$\\ $\gamma_2=-0.13^{+0.19}_{-0.20}$}  \\
			\hline
			this work & \makecell{158SGLS+JLA SNe Ia\\158SGLS+1598quasars}  & \makecell{$\Delta d/d\leq 5\%$for \\$\Lambda$CDM, (92)\\$\Delta d/d\leq 5\%$for\\ $\Lambda$CDM, (109(88))}  & \makecell{Maximum likelihood\\\& model independent}&\makecell{$\gamma_1=2.058^{+0.041}_{-0.040}$\\ $\gamma_2=-0.136^{+0.163}_{-0.165}$\\$\gamma_1=2.051^{+0.076}_{-0.077}$\\ $\gamma_2=-0.171^{+0.214}_{-0.196}$} \\
			\hline
			
	\end{tabular}}\label{tab3}
\end{table*}

\section{DISCUSSION AND CONCLUSIONS} \label{sec:DISCUSSIONS AND CONCLUSIONS}
In this paper, we adopt a model-independent approach for the first time to investigate whether the mass density power law index of SGLS evolves with redshift, using the sample of JLA SNe Ia  and the quasar sample from \cite{risa19} to provide the luminosity distances, and a relatively reasonable data pairing method to fit both the parameters of the luminosity distances and the index $\gamma$. Our work is based on the flat universe hypothesis and the CDDR and employs \textbf{Akaike weights and the BIC selection weights} to discriminate an evolutionary $\gamma$  from a nonevolutionary case. \textbf{We find that although the values of $\Delta AIC$ and $\Delta BIC$ cannot exclude either the evolutionary model or non-evolutionary model, with the help of the Akaike weights and the BIC selection weights, we can infer that the non-evolutionary model might have a very slight advantage, especially in relatively low redshift samples. This slight advantage becomes weaker with the addition of higher redshift data samples.} This means that an SIS model is well motivated for an SGLS sample with a relatively low redshift for $z_l<\sim 0.66$. \textbf{For high redshifts, an evolutionary $\gamma$ also fits the observed data well.} If the evolutionary case is considered, the $\gamma$ index decreases with redshift.\textbf{ Our results show that the $\gamma$ index is slightly more likely to be a non-evolutionary model in the case of the currently used samples,} however, this preference is less secure at higher redshifts. Larger and more accurate data samples may be needed in the future to distinguish between these two cases. We, therefore, suggest using SGLS models with the freedom of an evolving $\gamma$ index at higher redshifts.

\acknowledgments
We thank the anonymous referee for constructive comments. This work is supported by Yunnan Youth Basic Research Projects 202001AU070013, and the Institute of Astronomy of Dali University.

\bibliographystyle{aasjournal}

\end{document}